%% file: main.tex
\documentclass[submission,copyright,creativecommons]{eptcs}
\providecommand{\event}{ICLP 2021} 
\usepackage{underscore} 
\usepackage{mathptmx}
\usepackage{amsfonts}
\usepackage{amssymb}

\usepackage[T1]{fontenc}

\usepackage{caption}
\usepackage{comment}

\usepackage{graphicx}

\usepackage{xcolor}

\usepackage{csquotes}

\usepackage{paralist}

\usepackage{url}

\input{commands}

\title{Geolog: Scalable Logic Programming\\on Spatial Data}

\author{Tobias Grubenmann and Jens Lehmann
\institute{SDA Research Group, Department of Computer Science, University of Bonn, Germany}
\email{\{grubenmann, jens.lehmann\}@cs.uni-bonn.de}
\institute{Fraunhofer IAIS, Dresden, Germany}
         \email{jens.lehmann@iais.fraunhofer.de}
         }

\def\titlerunning{Geolog: Scalable Logic Programming on Spatial Data}
\def\authorrunning{T. Grubenmann \& J. Lehmann}

\begin{document}

\label{firstpage}

\maketitle

\begin{abstract}
\input{abstract}
\end{abstract}



\section{Introduction}

\input{intro}

\section{Preliminaries}

\input{preliminaries}

\section{{\relationEvaluation} Paradigm}

\input{strategies}

\section{Programming with the {\relationEvaluation} Strategy}\label{sec:programming}

\input{programming}

\section{The {\geolog} Tool}

\input{tool}

\section{Evaluation}\label{sec:evaluation}

\input{evaluation}

\section{Related Work}

\input{relatedwork}

\section{Conclusion}

\input{conclusion}

\section*{Acknowledgements}
\begin{sloppypar}
This work was partially funded by the Federal Ministry of Education and Research (BMBF), Germany under Simple-ML (01IS18054) and the European Commission under PLATOON (872592) and Cleopatra (812997).
Map data copyrighted OpenStreetMap contributors and available from \url{https://www.openstreetmap.org}.

Special thanks to G\"unter Kniesel-W\"unsche for his valuable insights into Logic Programming.
\end{sloppypar}

\bibliographystyle{eptcs}
\bibliography{bibliography}

\label{lastpage}
\end{document}

%% file: commands.tex
\newtheorem{definition}{Definition}
\newtheorem{example}{Example}

\newcommand{\entityEvaluation}{Entity Based Programming}
\newcommand{\relationEvaluation}{Relation Based Programming}

\newcommand{\geolog}{Geolog}

%% file: abstract.tex
Spatial data is ubiquitous in our data-driven society.
The Logic Programming community has been investigating the use of spatial data in different settings.
Despite the success of this research, the Geographic Information System (GIS) community has rarely made use of these new approaches.
This has mainly two reasons.
First, there is a lack of tools that tightly integrate logical reasoning into state-of-the-art GIS software.
Second, the scalability of solutions has often not been tested and hence, some solutions might work on toy examples but do not scale well to real-world settings.
The two main contributions of this paper are (1) the {\relationEvaluation} paradigm, expressing rules on relations instead of individual entities, and (2) {\geolog}, a tool for spatio-logical reasoning that can be installed on top of ArcMap, which is an industry standard GIS.
We evaluate our new {\relationEvaluation} paradigm in four real-world scenarios and show that up to two orders of magnitude in performance gain can be achieved compared to the prevalent {\entityEvaluation} paradigm.


%% file: intro.tex
Spatial data describes entities that have a \emph{shape}, in addition to other attributes.
A shape is a point, line, or polygon in a two- or three-dimensional space.
We call entities that have a shape \emph{spatial entities}.
What makes spatial data unique is that many relationships between the spatial entities are implicitly given by their shape.
On the one hand, this means that spatial data is very rich in implicit information.
For example, given two points, the distance between those points does not need to be stored explicitly in a data structure but instead, it is sufficient to store the coordinates of these points.
The distance can be computed whenever it is needed.
On the other hand, the implicit nature of most spatial relations also means that these relationships have to be computed on-the-fly, which can potentially require substantial computational resources.
We call the process of eliciting implicit relationships between spatial entities \emph{spatial reasoning}.

For more complex spatial relationships, spatial reasoning alone might not be enough to infer the knowledge required to answer a question.
In such cases, logical reasoning and spatial reasoning can be combined to answer the question.
We call these interweaving of logical and spatial reasoning \emph{spatio-logical reasoning}.
 
Spatio-logical reasoning has been popular in the last two decades.
For example, in the Semantic Web community, different solutions for reasoning over spatial data have been proposed~\cite{battleGeoSPARQLEnablingGeospatial2011,kyzirakosStrabonSemanticGeospatial2012,stockerPelletspatialHybridRCC82009}.
Meanwhile, in the Logic Programming (LP) community, spatial extensions for LP~\cite{hageSpacePackageTight2010,liLPrologQSFunctional2019} have been explored.

What all these approaches have in common is that spatial entities are represented as individual objects and relationships are established between these objects.
This is a very natural way of modelling spatial entities in a reasoning system, as the rules governing the relationships between them can be expressed on the level of individual spatial entities.
We call this approach the \emph{{\entityEvaluation}} paradigm.
This approach works especially well when there is only a single spatial operation required.
However, subsequent spatial operations have to be performed for each resulting object of the first spatial operation.

There is, however, an alternative way to model spatial entities by using relations. 
Many spatial operations can work with relations both as input and output. 
Therefore, we propose a \emph{{\relationEvaluation}} paradigm, in which spatial operations create relations as outputs, which can be used as input for subsequent spatial operations. 
Hence, our research hypothesis is that \emph{passing references to relations is more efficient than iterating over individual objects}.
Indeed, as we show in the evaluation section, this is the case for different scenarios.
Thus, we propose to drop the restriction that reasoning rules on spatial entities should be expressed on individual entities and instead, we show that for scalability reasons, it is often more efficient to express certain reasoning rules on relations.
Hence, the problem statement of this work can be described as:
\emph{Can the scalability of spatio-logical reasoning be improved by expressing reasoning rules on relations instead of entities?}

In this work, we present 
\begin{inparaenum}[(1)]
    \item the \emph{{\relationEvaluation}} paradigm, which expresses reasoning rules on spatial data using relations,
    \item \emph{{\geolog}}, a new reasoning plugin for ArcMap based on Logic Programming enabling {\relationEvaluation} in addition to {\entityEvaluation}, and
    \item a comparison of the scalability of the different programming paradigms.
\end{inparaenum}
ArcMap (a.k.a ArcGIS Desktop) is a GIS application that is considered an industry standard~\cite{khanEmpiricalEvaluationArcGIS2017}.  
At its core, {\geolog} provides the ability to use the ArcMap API within Prolog.
For this, {\geolog} integrates the arcpy library---a Python API for ArcMap---and PySwip---a Python interface to SWI Prolog---and thus, provides the ability to use many spatial functions that are provided by ArcMap.

\begin{figure}[htb]
    \centering
    \includegraphics[width=\linewidth, trim=0 0 0 0, clip]{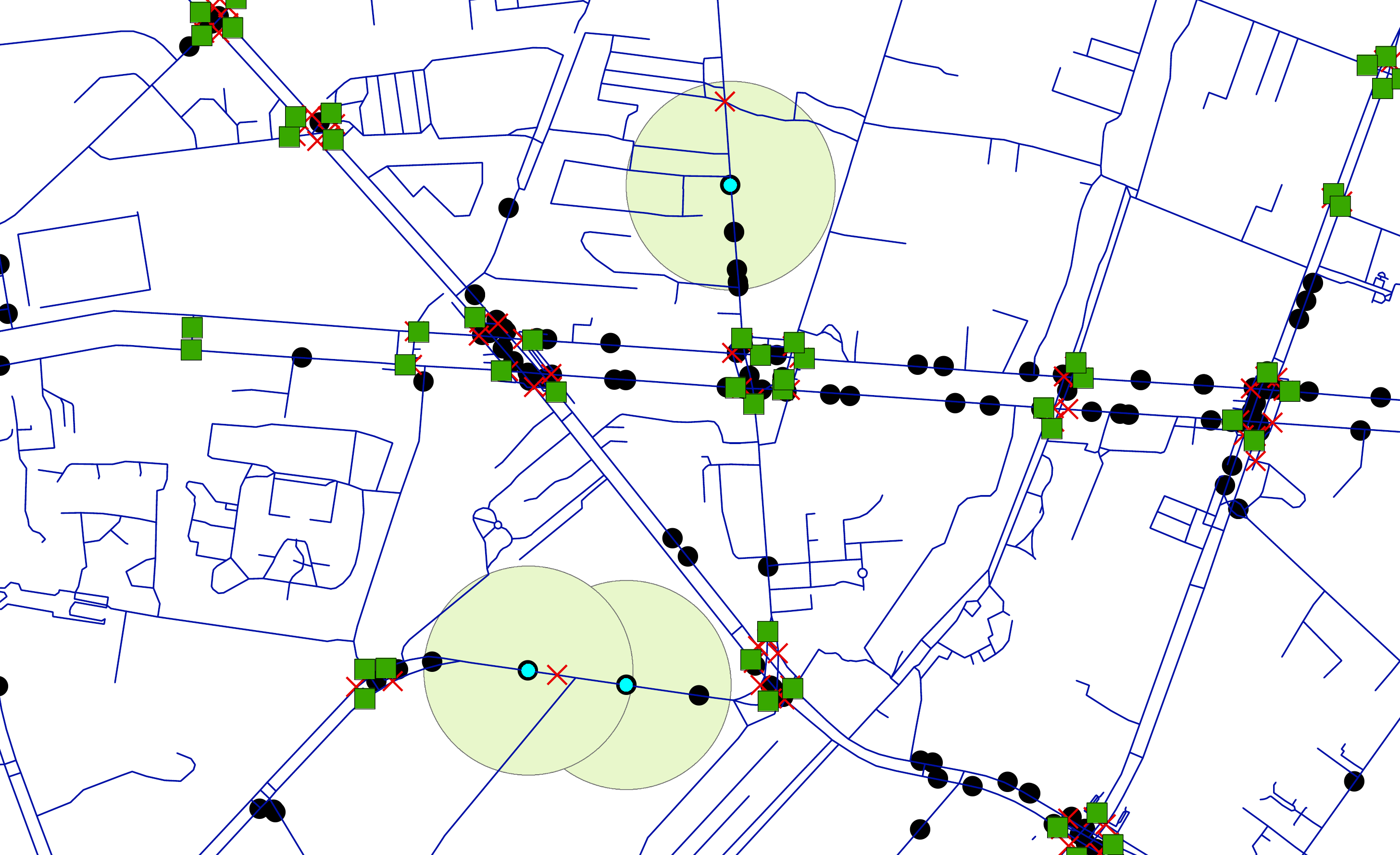}
    \caption{Selected accidents ($\circ$) near crossings ($\times$) without traffic lights ($\blacksquare$), and unselected accidents ($\bullet$)}
    \label{fig:arcMapQ4Result}
\end{figure}

Finally, to enable GIS users to make use of the reasoning power of any reasoning system, it is important to provide the means to integrate the reasoner into existing GIS software and GIS workflows.
Spatial data is visual by nature and hence, it is desirable that the output of the reasoning process can be seamlessly integrated into GIS software that is designed to visualize and further process the spatial data.
{\geolog} achieves this via side-effects of certain predicates.
For example, one side-effect could be that a predicate creates a new selection within an ArcMap map document (Figure~\ref{fig:arcMapQ4Result}).
Such a new selection allows the GIS user to directly continue with these newly selected entities within or outside of {\geolog}.
We implement {\geolog} as an ArcMap Python-Addin which can be easily installed on the most recent versions of ArcMap\footnote{At the time of writing, the most recent version is ArcMap 10.8.1}.
Full integration is achieved by incorporating the arcpy library into {\geolog}.




%% file: preliminaries.tex
\emph{Spatial operations} take spatial data as input.
Spatial data is any kind of data that has a \emph{shape} (point, line string, polygon) attached to its entities.
We call such entities \emph{spatial entities}.
For example, determining the distance between two Points of Interest (POIs) on a map is a spatial operation taking two spatial entities as input.
Besides other information like the name of the POI or the category, each POI also has a shape, in this case a point.
The spatial operation uses these shapes to define the output, i.e., the distance between the two points defining the location of the POIs.
Spatial data can be very rich in information because many relationships are implicitly  given by the location and orientation of different shapes.
\emph{Spatial reasoning} can be seen as the task of making these implicitly  given relationships explicit.

A \emph{spatial predicate} is a Prolog-predicate which makes use of spatial operations to determine the truth-value of the predicate or to instantiate unbound variables.
For instance, let \texttt{distance(X, Y, D)} be a predicate which is true when $D$ is the distance between two points $X$ and $Y$.
In this case, \texttt{distance(X, Y, D)} is a spatial predicate.

Spatio-logical reasoning incorporates logical reasoning on spatial and non-spatial entities and spatial reasoning on spatial entities.
%
Whereas both, logical and spatial reasoning, have the same goal of making implicit facts explicit, spatial reasoning is often performed in dedicated libraries or systems.
The reason for this is that
\begin{inparaenum}[(1)]
    \item calculating spatial relationships is a non-trivial task requiring dedicated algorithms,
    \item different coordinate systems and projections require reprojecting, and
    \item spatial indices might be required for scalable solutions.
\end{inparaenum}
However, having a dedicated library can also pose some challenges, as the calls to this library need to be interwoven into the logical reasoning process. The remainder of this section will discuss the two different approaches using the example below.

\begin{example}\label{exp:sameStreet}
    The goal is to find for each accident the nearby traffic entities (signals, crossings, etc.) and the streets to which these traffic entities belong. $\blacktriangle$
\end{example}

To understand the difference between the programming paradigms mentioned in this work, we need to distinguish between a \emph{logical layer} and a \emph{spatial layer}.
The logical layer is the part of the reasoning process that uses logical reasoning to establish new facts.
In our specific implementation, the logical layer corresponds to all reasoning steps performed within Prolog.
The spatial layer performs spatial operations using the underlying spatial libraries (or spatial databases).
A \emph{spatial predicate} inside the logical layer is a predicate that uses a spatial operation, directly or indirectly, in its definition.
Note that a spatial predicate can use more than one spatial operation.

\subsection*{{\entityEvaluation} Paradigm}

The \emph{{\entityEvaluation}} paradigm is as follows:
Each spatial entity is modelled as a separate atom in the logical layer.
If a call to the spatial operation has no unbound variables, the spatial operation returns true or false.
If a call to the spatial operation has unbound variables, the spatial operation returns the atoms corresponding to all spatial entities matching the spatial operation in an iterative manner.

To improve the performance of the {\entityEvaluation} paradigm, the spatial entities can be stored in a backend storage (e.g., a spatially enabled database \cite{kyzirakosStrabonSemanticGeospatial2012}).
Using such a backend, it is possible to exploit spatial indices to quickly elicit the result set of all matching spatial entities.
Once all spatial entities are found, the individual atoms are returned by iterating over the result set and matching the entity to the corresponding atom.

We will now discuss how the {\entityEvaluation} paradigm is used in Example \ref{exp:sameStreet}.
For this, we assume that there is a spatial predicate \texttt{close(X, Y)} which is true if the spatial entity corresponding to atom \texttt{X} is not farther away than 10 metres from the spatial entity corresponding to atom \texttt{Y}.
Using \texttt{close}, we can define that an entity \texttt{X} is on a street \texttt{Y}, if they are closer than 10 metres.
%
We further assume that there are predicates \texttt{accident(X)}, \texttt{traffic(X)}, and \texttt{street(X)}, which are true when \texttt{X} is an atom representing an accident, a traffic feature, and a street, respectively.
Finally, we assume a predicate \texttt{near(X, Y)} which is true if a spatial entity denoted by atom \texttt{X} is not farther away than 100 metres from a spatial entity denoted by atom \texttt{Y}.

A possible query for this example could look like the one in Figure \ref{lst:entityEvaluationExample}.

\begin{figure}
\begin{verbatim}
nearbyTrafficAndStreet(A, T, S) :- accident(A), near(A, T), traffic(T),
                                   close(T, S), street(S).
\end{verbatim}
\caption{A predicate for the {\entityEvaluation} paradigm for traffic entities near accidents and their street.}\label{lst:entityEvaluationExample}
\end{figure}


\begin{figure}
\begin{verbatim}
accident(a1). traffic(t1). traffic(t2). street(s1). street(s2).
\end{verbatim}
\caption{Facts assumed for the predicate in Figure \ref{lst:entityEvaluationExample}.}\label{lst:entityEvaluationFacts}
\end{figure}

Let us further assume the facts in Figure \ref{lst:entityEvaluationFacts}.
Let us assume that \texttt{t1} and \texttt{t2} are near \texttt{a1}, and that \texttt{s1} is close to \texttt{t1} and \texttt{s2} close to \texttt{t2}, then the predicate in Figure \ref{lst:entityEvaluationExample} is evaluated as follows.
First, accident \texttt{a1} is retrieved on the logical layer using predicate \texttt{accident(A)}.
Then, the spatial predicate is used to determine which spatial entities \texttt{T} are near \texttt{a1}.
For this, a spatial operation is used which determines the result set \{\texttt{t1}, \texttt{t2}\}.
The spatial operation returns first \texttt{t1}, which is tested by predicate \texttt{traffic(T)} whether it is a traffic entity or not.
As it is the case, the spatial predicate \texttt{close(T, S)} is used next to determine the entities \texttt{S} that are close to \texttt{t1}.
The result set of this second spatial operation is \{\texttt{s1}\} and hence, the spatial predicate returns \texttt{s1} as the only match.
Finally, \texttt{s1} is tested whether it is a street or not.
As it is the case, we have found the first solution for the predicate \texttt{nearbyTrafficEntityOnStreet(A, T, S)}: \texttt{\{A: a1, T: t1, S: s1\}}.
Now, we can backtrack and retrieve the second entity matching the predicate \texttt{near(a1, T)}.
The same steps are performed on \texttt{t2} to obtain the second solution: \texttt{\{A: a1, T: t2, S: s2\}}.

The important observation here is that each matching traffic entity for \texttt{near(a1, T)} results in a separate call to the predicate \texttt{close(T, S)}.
Also, for every call we need to send atoms corresponding to spatial entities back and forth between the logical and spatial layer.


%% file: strategies.tex
In the following, we introduce our new {\relationEvaluation} Paradigm, which complements the {\entityEvaluation} Paradigm.
Note that both programming paradigms can be mixed within a single Logic Program by, either, iterating over a relation to get single entities from a relation, or, collecting single entities and store them in a new relation.

The \emph{{\relationEvaluation}} paradigm is as follows:
\emph{Relations} are used to represent sets of entities and relationships between entities.
These relations are represented on the logical layer as specific atoms, where a single atom in the logical layer can correspond to a relation consisting of an arbitrary number of entities.
Correspondingly, the spatial predicates take as input atoms representing those relations rather than individual entities.
In practice, such relations can be realized as database relations.
We will denote relations which store a collection of entities as \emph{entity-relation} and relations which store relationships between different entities as \emph{relationship-relation}. 
Note that entity-relations contain all the attributes associated to a spatial entity, including its shape.
In contrast, a relationship-relation only contains the foreign keys which point to the entities in an entity-relation for which the relationship is true.

To understand how the \emph{{\relationEvaluation}} paradigm works, we come back to Example~\ref{exp:sameStreet}.
Figure~\ref{lst:relationEvaluationExample} illustrates how the predicate for Example~\ref{exp:sameStreet} could be defined.
Most notably, the predicate has now a fourth parameter $R$ which represents the relationship-relation that will be produced by the spatial operation.
As a first step, $A$, $T$, and $S$ will be bound to the entity-relations containing all accidents, traffic entities, and streets, respectively.
Afterwards, the \texttt{near(A, T, R1)} relation binds $R1$ to the relationship-relation of all pairs of accidents that are near traffic entities.
Next, the \texttt{close(T, S, R2)} relation binds $R2$ to the relationship-relation of all pairs of traffic entities that are close to streets.
Finally, a join between $R1$ and $R2$ yields the relationship-relation $R$ that contains all triples of accidents, traffic entities, and streets such that the traffic entity is near the accident and the street is close to the traffic entity.

\begin{figure}
\begin{verbatim}
nearbyTrafficAndStreet(A, T, S, R) :- accidents(A), traffics(T), streets(S),
                                      near(A, T, R1), close(T, S, R2),
                                      join(R1, R2, R).
\end{verbatim}
\caption{A predicate for the {\relationEvaluation} paradigm for traffic entities near accidents and their street.}\label{lst:relationEvaluationExample}
\end{figure}

%% file: programming.tex
In the following, we discuss the different predicates that we use in the evaluation section for both the {\entityEvaluation} paradigm and {\relationEvaluation} paradigm.

For the {\relationEvaluation} paradigm, we distinguish between \emph{entity-relations}, which contain spatial entities with the associated attributes including the shape, and \emph{relationship-relations}, which contain tuples of foreign keys but no shape.

Entities are uniquely identified by a pair of the form \texttt{(category, ID)}, e.g., \texttt{("accidents", 1)}.
Relations are uniquely identified by the name of the relation, e.g., \texttt{"accidents"}.

\subsection{Spatial Predicates}

Spatial predicates are naturally an important part of spatio-logical reasoning.
For the {\entityEvaluation} paradigm, spatial predicates are defined the following way:
\begin{definition}[Entity-based near predicate]
    The predicate \texttt{near(E\_1, E\_2)} is true if the distance between the entities \texttt{E\_1} and \texttt{E\_2} is not larger than 100 metres. $\blacktriangle$
\end{definition}
The predicate \texttt{closeby(E\_1, E\_2)} is defined analogously, with a threshold of 10 metres instead of 100 metres.

For the {\relationEvaluation} paradigm, spatial predicates are defined the following way:
\begin{definition}[Relation-based near predicate]
    The predicate \texttt{near\_relational(E\_1, E\_2, R, [A\_1, A\_2])} is true if \texttt{R} is a relationship-relation whith attributes \texttt{A\_1} and \texttt{A\_2} which contain all keys from entity-relations \texttt{E\_1} and \texttt{E\_2}, respectively, such that the spatial entities corresponding to those keys are not farther away than 100 metres. $\blacktriangle$
\end{definition}

The predicate \texttt{closeby\_relational(E\_1, E\_2, R, [A\_1, A\_2])} is defined analogue, with a threshold of 10 metres instead of 100 metres.

Given such a relation-based spatial predicate, it is often useful to restrict the further reasoning process on those entities that satisfy a given spatial predicate.
We can use the identifiers in the relationship-relation to filter for those points:
\begin{definition}[Relation-based filter predicate]
    The predicate \texttt{filter\_by\_relationship(E\_IN, R, A, E\_OUT)} is true if the entity-relation \texttt{E\_OUT} contains all spatial entities which are in entity-relation \texttt{E\_IN} and their key is contained in the attribute \texttt{A} of relationship-relation \texttt{R}. $\blacktriangle$
\end{definition}

\subsection{OSM-Specific Predicates}

The following predicates are specific for identifying different types of entities, as defined by OpenStreetMap (OSM)~\cite{openstreetmapcontributorsPlanetDumpRetrieved2017}.
OSM uses a 4-digit code to determine the type of a point, line, or polygon.
In addition to specific types, OSM also defines code ranges for supertypes.
E.g., schools have the code 2082, whereas entities related to education have an OSM code between 2080 and 2089.
Using these codes, it is possible to define rules which apply only to certain types of entities.
The following two predicates are used to determine or test the type of an entity.

\begin{definition}[Entity-based type predicate]
    The predicate \texttt{entity\_type(T, E)} is true if entity \texttt{E} is of type \texttt{T}. $\blacktriangle$
\end{definition}

\begin{definition}[Relation-based type predicate]
    The predicate \texttt{entity\_type\_relational(T, E\_IN, E\_OUT)} is true if entity-relation \texttt{E\_OUT} contains all entities in entity-relation \texttt{E\_IN} which are of type \texttt{T}. $\blacktriangle$
\end{definition}

\subsection{Relational-Specific Predicates}

The following three predicates are only used when working with relations.

\begin{definition}[Relation-based join predicate]
    The predicate \texttt{join\_relational(E\_1, E\_2, E\_OUT, A)} is true if the relationship-relation \texttt{E\_OUT} contains all pairs of entities from relationship-relations \texttt{E\_1} and \texttt{E\_2} which have the same value on attribute \texttt{A}. $\blacktriangle$
\end{definition}

The projection can be helpful to remove attributes from a relationship-relation which are not needed anymore.

\begin{definition}[Relation-based projection]
    The predicate \texttt{project\_id\_relational(E\_in, L, E\_OUT)} is true if relationship-relation \texttt{E\_OUT} contains all attributes in list \texttt{L} for each record in relationship-relation \texttt{E\_IN}. $\blacktriangle$
\end{definition}

Finally, the relation-based difference is defined as follows:

\begin{definition}[Relation-based difference]
    The predicate \texttt{minus\_relational(E\_1, E\_2, E\_OUT)} is true if relationship-relation \texttt{E\_OUT} contains all entities in relationship-relation \texttt{E\_1} that are not in relationship-relation \texttt{E\_2}. \texttt{E\_1} and \texttt{E\_2} must have the same attributes. $\blacktriangle$
\end{definition}

%% file: tool.tex
\begin{figure}[htb]
	\centering
	\includegraphics[width=0.7\textwidth, trim=220 95 95 110, clip]{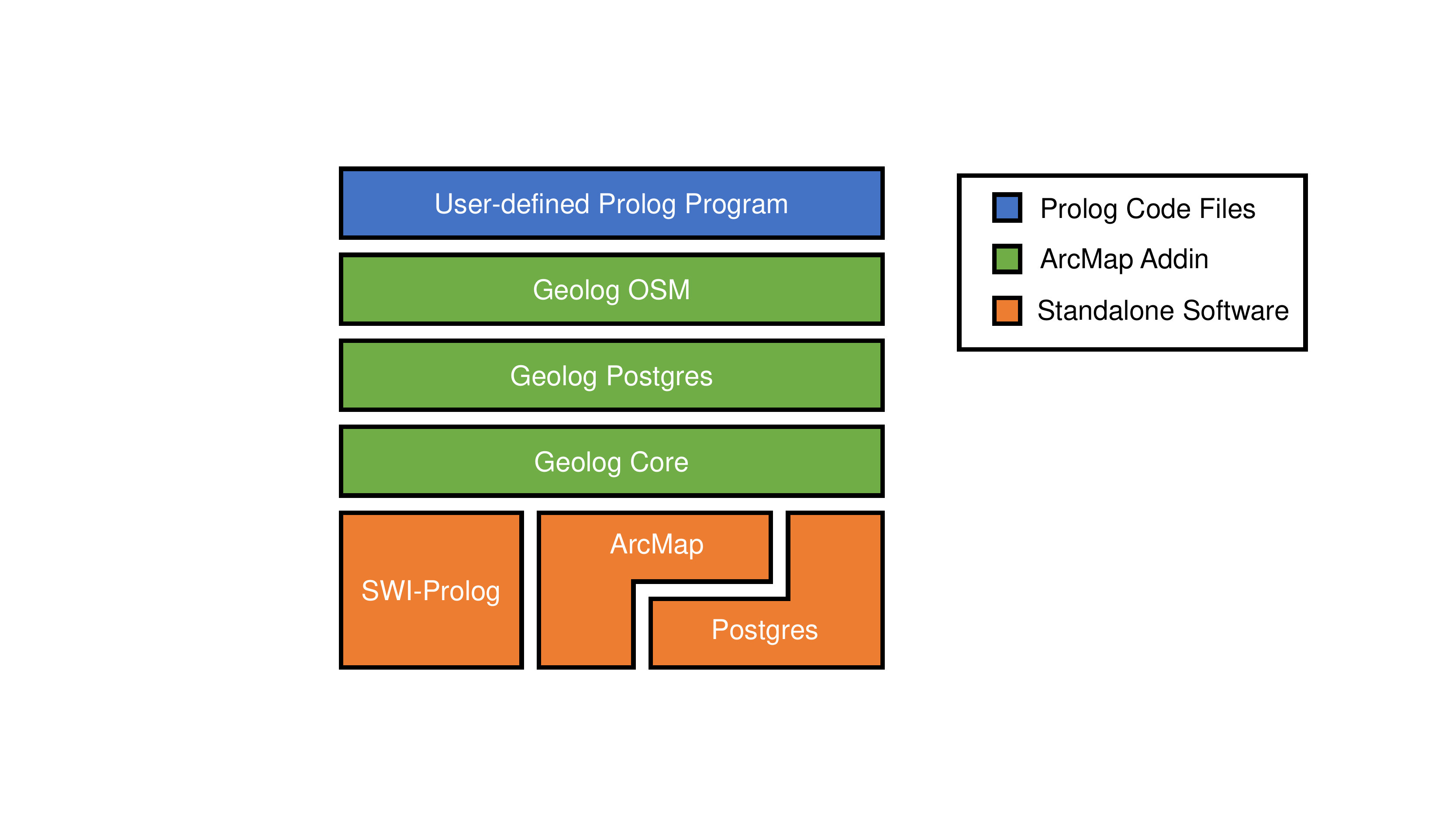}
	\caption{The different layers of {\geolog}.}
	\label{fig:geologOverview}
\end{figure}

In Figure \ref{fig:geologOverview}, we illustrate the different layers of our {\geolog} approach.
At the bottom, in orange, we have SWI-Prolog, ArcMap, and Postgres as standalone software components.
Note that the access to Postgres can be via ArcMap or directly from the {\geolog} core.

Above the standalone software, in green, is the \emph{ArcMap Addin}, which consists of three different components.
Addins are a plugin concept for ArcMap and allow easy installation of tools and toolbars with simple UI elements.
The first component, \emph{Geolog Core}, provides core functionalities which are required to communicate with ArcMap.
This component also handles communication with the Prolog interpreter and keeps track of which Prolog atom corresponds to which Python object.
The next component, \emph{Geolog Postgres}, provides predicates which facilitates the communication with the database.
All predicates from Section \ref{sec:programming}, except the OSM specific predicates, are implemented in this component.
Finally, the \emph{Geolog OSM} component provides predicates for filtering different OSM types and the atoms corresponding to these types.

The top most layer, in blue, contains the application-specific Prolog code written by the user.
In Section \ref{sec:evaluation}, we introduce different scenarios with some examples of user code.

To communicate with ArcMap, the Geolog Core Addin uses the \emph{arcpy} library, which provides a Python interface for most functionalities within ArcMap.
Geolog Core maps 398 classes and 3058 functions from the arcpy library and makes them available to a Prolog Programm.\footnote{The reported number of classes and functions is for ArcMap 10.8.1 with Advanced License. The number might differ depending on the version and license.}
These classes and functions not only provide results that can be further processed within {\geolog} but they often also have side-effects which can be exploited.
For example, it is possible to manipulate a map object using arcpy to illustrate the result of the program or to create data structures for further processing.
In Figure \ref{fig:arcMapQ4Result}, we have created a map that illustrates all accidents near crossings without traffic lights.
For this, we extended the program from Figure \ref{lst:relationQ4} to draw a 100 meter radius around each accident (to illustrate which objects are near), to create a new collection with all crossings, to create a new collection with all traffic lights, and to select all accidents that match our criteria (near a crossing that has no traffic light).

%% file: evaluation.tex
\begin{figure}[htb]
    \centering
    \begin{minipage}{.47\textwidth}
        \centering
        \includegraphics[width=\linewidth, trim=70 280 65 270, clip]{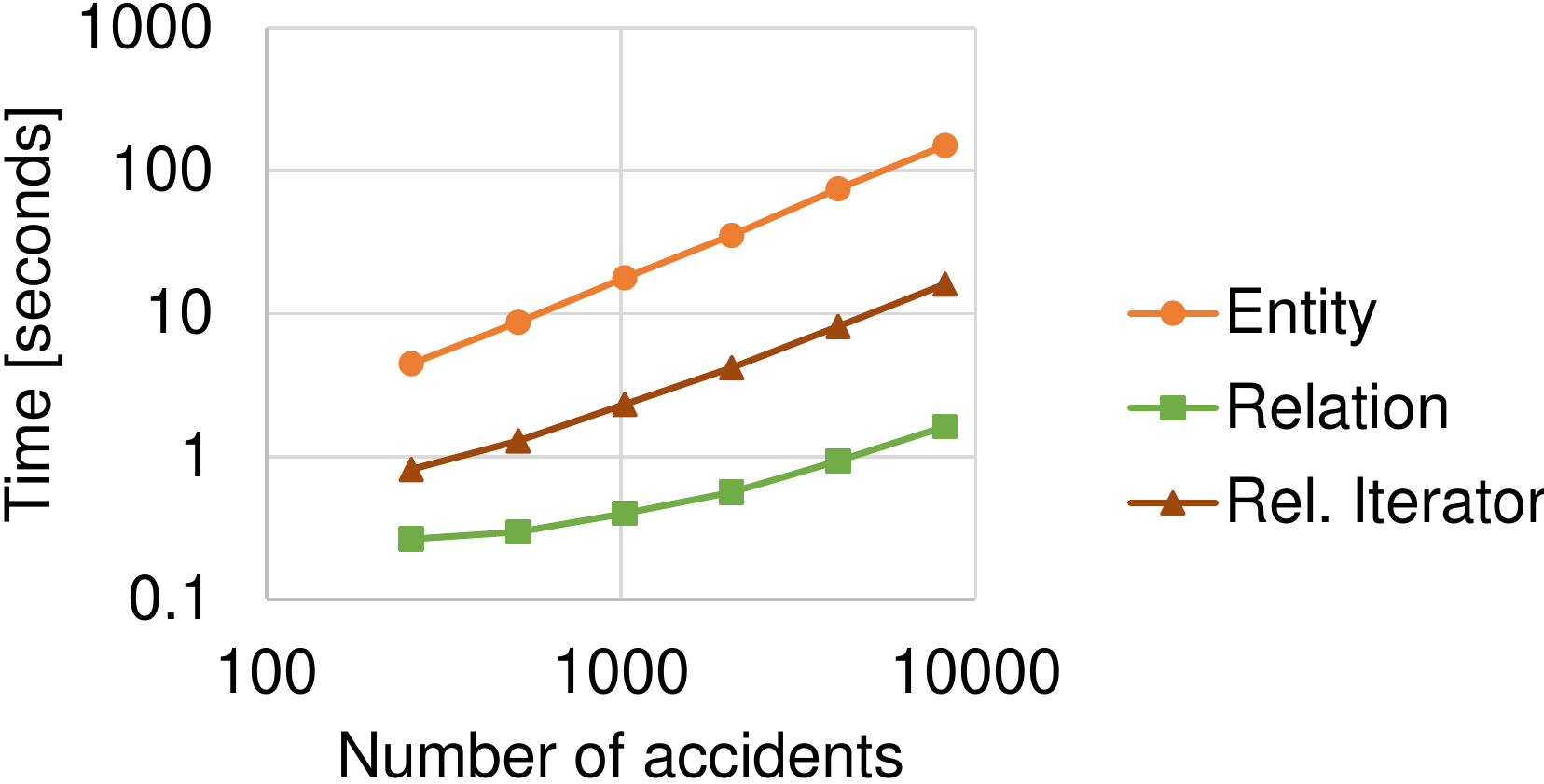}
        \caption{Scaling for Scenario 1}
        \label{fig:scalingQ1}
    \end{minipage}\qquad%
    \begin{minipage}{.47\textwidth}
        \centering
        \includegraphics[width=\linewidth, trim=70 280 65 270, clip]{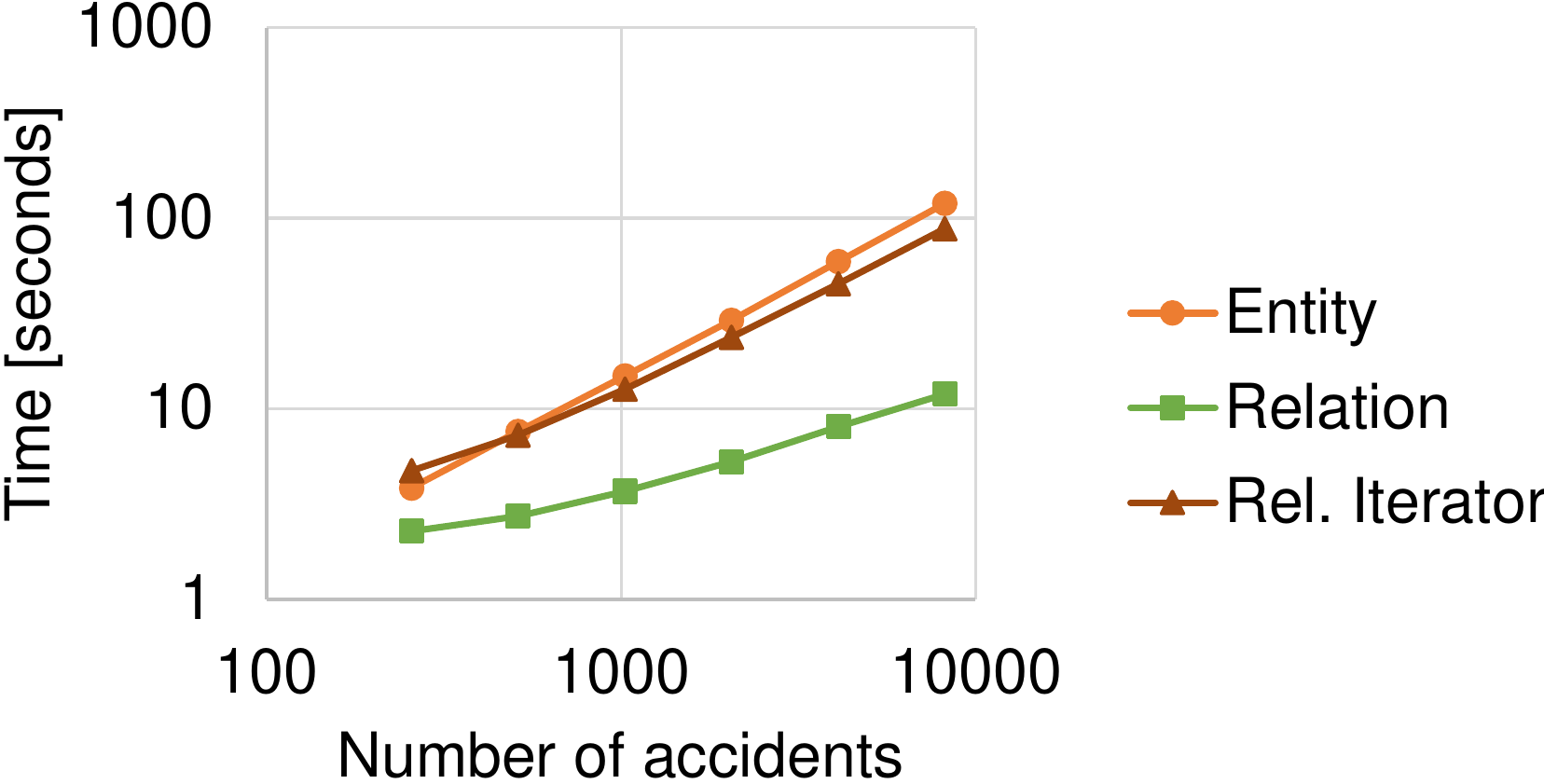}
        \caption{Scaling Scenario 2}
        \label{fig:scalingQ2}
    \end{minipage}

    \centering
    \begin{minipage}{.47\textwidth}
        \centering
        \includegraphics[width=\linewidth, trim=70 280 65 240, clip]{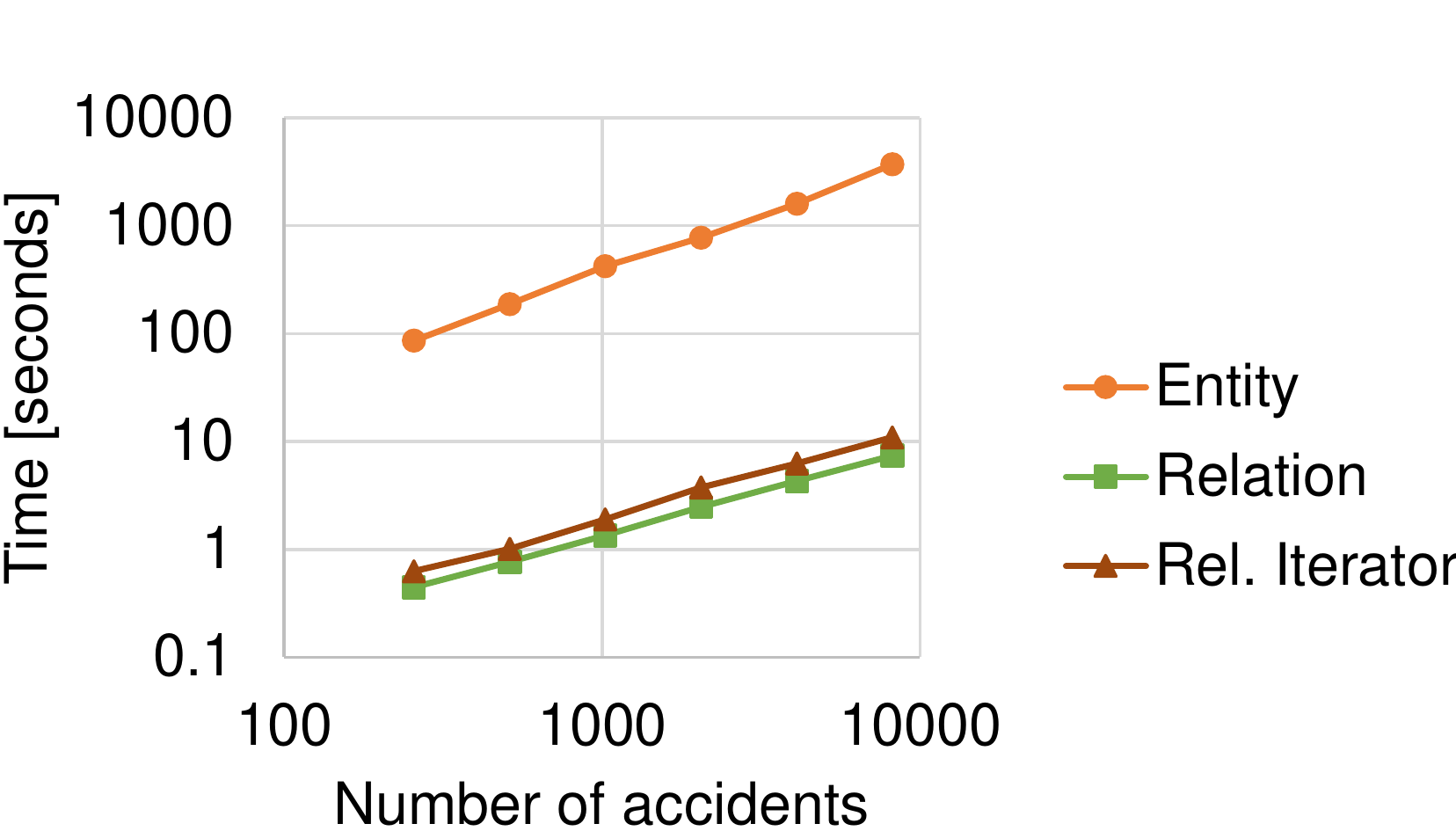}
        \caption{Scaling Scenario 3}
        \label{fig:scalingQ3}
    \end{minipage}\qquad%
    \begin{minipage}{.47\textwidth}
        \centering
        \includegraphics[width=\linewidth, trim=70 280 65 240, clip]{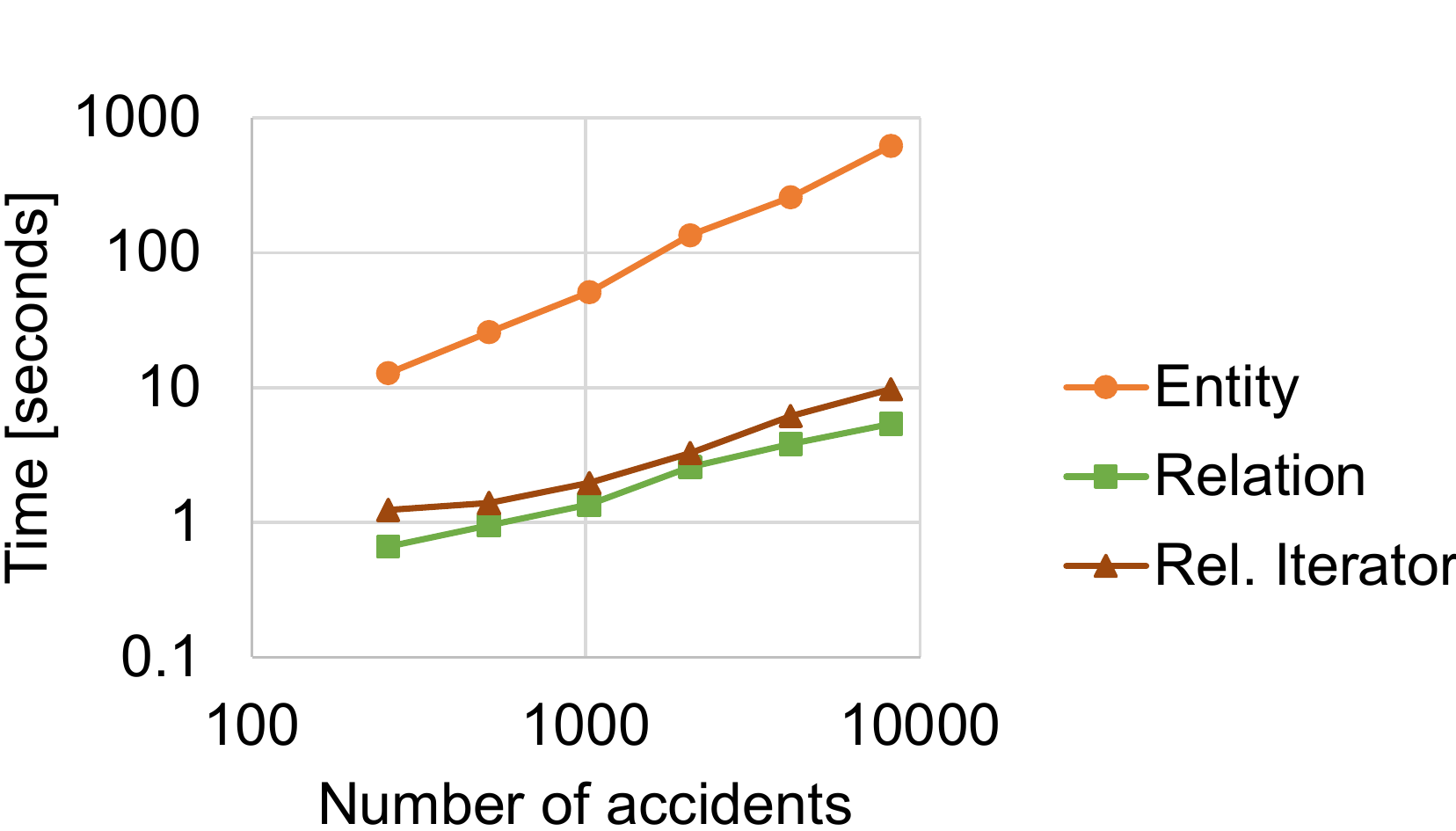}
        \caption{Scaling for Scenario 4}
        \label{fig:scalingQ4}
    \end{minipage}
\end{figure}

In the following, we discuss the scaling behaviour of four different scenarios using both the {\entityEvaluation} (\textsf{Entity} in Figures \ref{fig:scalingQ1}--\ref{fig:scalingQ4}) and {\relationEvaluation} (\textsf{Relation} in Figures \ref{fig:scalingQ1}--\ref{fig:scalingQ4}) strategies.
In addition, we also evaluate the scaling behaviour when there is a need to iterate over the result of the {\relationEvaluation} paradigm and thus, adding an additional iteration over the result at the end of the query (\textsf{Rel. Iterator} in Figures \ref{fig:scalingQ1}--\ref{fig:scalingQ4}).
The scaling is evaluated with respect to the number of accidents.
For this, we randomly sample a certain number of accidents from all  accidents in Berlin.
Each point in the plot is the average of ten runs.
All scaling plots are in log-log scale.

The scenarios focus on the analysis of accidents in Berlin, Germany.
For this, we use the official accident data of 2019.\footnote{\url{https://unfallatlas.statistikportal.de/\_opendata2020.html}}
To establish the context of the accidents, we use the OpenStreetMap data for Berlin, prepared by Geofabrik.\footnote{\url{http://download.geofabrik.de/}}
For the \texttt{near} predicates, a threshold of 100 metres has been chosen and a threshold of 10 metres for the \texttt{closeby} predicates.

All experiments were conducted on a PC with 16GB of RAM and a 1.6GHz 4 core processor.
As software, we used ArcMap 10.8.1 (32-bit), Python 2.7.18 (32-bit), SWI-Prolog 8.2.4 (32 bit), and the SWI-Prolog-Python interface PySwip 0.2.10\footnote{\url{https://github.com/yuce/pyswip}}.

For the evaluation, v0.1.1 of the Geolog Addin was used, available at \url{https://github.com/tobiasgrubenmann/geolog_addin}.
The Prolog code for the evaluation is available at \url{https://github.com/tobiasgrubenmann/geolog_accident}.

\subsection{Scenario 1: Accidents near crossings}

The aim of this first scenario is to establish how the different programming paradigms perform on a simple query with one spatial operation.
For this, we query for accidents near pedestrian crossings.
The two queries for the {\entityEvaluation} (Figure \ref{lst:entityQ1}) and {\relationEvaluation} (Figure \ref{lst:relationQ1}) both use one spatial predicate and one type predicate.
The scaling is shown in Figure \ref{fig:scalingQ1}. 
As we can see from the figure, all three approaches have the same scaling behaviour.
However, the execution times are roughly one order of magnitude apart from each other.
For example, to analyse 8192 accidents, the {\entityEvaluation} requires around 150 seconds, the {\relationEvaluation} with an iterator at the end requires only 16 seconds, and the {\relationEvaluation} without iterator requires only 1.6 seconds for the same query.

\begin{figure}
\begin{verbatim}
:- near(("accidents", AccidentID), ("traffic", TrafficID)),
   entity_type(crossing_features, ("traffic", TrafficID)).
\end{verbatim}
\caption{Query for Scenario 1 using the {\entityEvaluation} paradigm}\label{lst:entityQ1}
\end{figure}

\begin{figure}
\begin{verbatim}
:- entity_type_relational(crossing_features, "traffic", Crossings),
   near_relational("accidents", Crossings, Result).
\end{verbatim}
\caption{Query for Scenario 1 using the {\relationEvaluation} paradigm}\label{lst:relationQ1}
\end{figure}

\subsection{Scenario 2: Traffic entities on the same street as accidents}

The aim of this scenario is to test the scaling behaviour when two spatial operations are involved which have to be chained after each other.
For this, we query for accidents and traffic features that are on the same road.
For the {\entityEvaluation}, we just chained two spatial predicates after each other (Figure \ref{lst:entityQ2}).
In contrast, the {\relationEvaluation} (Figure \ref{lst:relationQ2}) has an additional filter predicate to restrict further processing to those roads that are close to accidents.
Also, the {\relationEvaluation} requires a final join to retrieve all accidents that share the same road as the traffic entities.

As one can see in Figure \ref{fig:scalingQ2}, \textsf{Entity} and \textsf{Rel.~Iterator} are close together.
This can be explained by the fact that the number of intermediate results (streets) is considerably smaller than the end result (traffic entities and their streets for each accident).
Therefore, the {\relationEvaluation} is only beneficial when there is no need to iterate over the large end result.
In contrast, \textsf{Relation} is one order of magnitude faster than \textsf{Entity} and in addition has a flatter scaling curve.

\begin{figure}
\begin{verbatim}
:- closeby(("accidents", AccidentID), ("roads", RoadID)),
   closeby(("traffic", TrafficID), ("roads", RoadID)).
\end{verbatim}
\caption{Query for Scenario 2 using the {\entityEvaluation} paradigm}\label{lst:entityQ2}
\end{figure}

\begin{figure}
\begin{verbatim}
:- closeby_relational("accidents", "roads", AccRoads, ["Acc", "Road"]),
   filter_by_relationship("roads", AccRoads, "Road", Roads),
   closeby_relational("traffic", Roads, TrafficRoads, ["Traffic", "Road"]),
   join_relational(AccRoads, TrafficRoads, Result, "Road",
      ["Acc", "rel1.Road", "Traffic"]).
\end{verbatim}
\caption{Query for Scenario 2 using the {\relationEvaluation} paradigm}\label{lst:relationQ2}
\end{figure}

\subsection{Scenario 3: Accidents near POIs that are near schools}

The third scenario queries for accidents that are near points of interest (POIs) that are near schools.
The scenario is similar to the last one, in the sense that both are chaining two spatial predicates.
However, this time, the intermediate results will be much larger since there are many POIs on the map.
Indeed, as Figure \ref{fig:scalingQ3} shows, \textsf{Rel. Iterator} is very close to \textsf{Relation}.
In addition, both outperform \textsf{Entity} by more than two orders of magnitude.
For example, for 8192 accidents, \textsf{Rel. Iterator} and \textsf{Relation} require around 10 seconds, whereas \textsf{Entity} requires almost an hour to get the same result.

The two queries for this scenario are shown in Figures~\ref{lst:entityQ3} and \ref{lst:relationQ3}.

\begin{figure}
\begin{verbatim}
:- near(("accidents", AccidentID), ("pois", Pois1)),
   near(("pois", Pois1), ("pois", Pois2)),
   entity_type(school_features, ("pois", Pois2)).
\end{verbatim}
\caption{Query for Scenario 3 using the {\entityEvaluation} paradigm}\label{lst:entityQ3}
\end{figure}

\begin{figure}
\begin{verbatim}
:- near_relational("accidents", "pois", AccidentPois, ["Acc", "Poi"]),
   filter_by_relationship("pois", AccidentPois, "poi", Pois),
   entity_type_relational(school_features, "pois", Schools),
   near_relational(Pois, Schools, PoisSchools, ["Poi", "School"]),
   join_relational(AccidentPois, PoisSchools, AccidentPoiSchool, "Poi",
      ["Acc", "rel1.Poi", "School"]).
\end{verbatim}
\caption{Query for Scenario 3 using the {\relationEvaluation} paradigm}\label{lst:relationQ3}
\end{figure}

\subsection{Scenario 4: Accidents near crossings without traffic lights}

The final scenarios are testing the scaling behaviour when querying with negation.
We are looking for accidents that happened at crossings that have no traffic lights nearby.
In the {\entityEvaluation} paradigm, we use the negation \texttt{\textbackslash+} to test for the absence of traffic lights (Figure \ref{lst:entityQ4}).
For the {\relationEvaluation} paradigm (Figure \ref{lst:relationQ4}), we make use of the \texttt{minus\_relation} predicate to substract the pairs of accidents and crossings with traffic lights from the relation containing all pairs of accidents and crossings. The results are pairs of accidents and crossings without traffic lights.
Since the \texttt{minus\_relation} predicate requires the schemas of both relations to match, we also use a projection predicate.

As one can see in Figure~\ref{fig:scalingQ4}, the scaling behaviour of \textsf{Relation} and \textsf{Rel. Iterator} are close.
This can be explained by the fact that the end result is quite small and hence, the overhead of iterating over the end result does not have such a big impact.
\textsf{Entity} is almost two orders of magnitude slower than the others for 8192 accidents.
Moreover, the slope of the scaling curve is steeper for \textsf{Entity}.

\begin{figure}
\begin{verbatim}
:- near(("accidents", AccidentID), ("traffic", Traffic)),
   entity_type(crossing_features, ("traffic", Traffic)),
   \+(near(("traffic", Traffic), ("traffic", OtherTraffic)),
      entity_type(traffic_signal_features, ("traffic", OtherTraffic))).
\end{verbatim}
\caption{Query for Scenario 4 using the {\entityEvaluation} paradigm}\label{lst:entityQ4}
\end{figure}

\begin{figure}
\begin{verbatim}
:- entity_type_relational(crossing_features, "traffic", Crossings),
   near_relational("accidents", Crossings, AccCrossing, ["Acc", "Crossing"]),
   filter_by_relationship("traffic", AccCrossing, "Crossing", CrossFilt),
   entity_type_relational(traffic_signal_features, "traffic", Signals),
   near_relational(CrossFilt, Signals, CrossingsSignals, ["Crossing", "Sig"]),
   join_relational(AccCrossing, CrossingsSignals, Join, "Crossing", 
      ["Acc", "rel1.Cross", "Sig"]),
   project_id_relational(Join, ["Acc", "Crossing"], AccSignal),
   project_id_relational(Accrossing, ["Acc", "Crossing"], AllAcc),
   minus_relational(AllAcc, AccSignal, Result).
\end{verbatim}
\caption{Query for Scenario 4 using the {\relationEvaluation} paradigm}\label{lst:relationQ4}
\end{figure}

%% file: relatedwork.tex
Region Connection Calculus (RCC) \cite{cohnQualitativeSpatialRepresentation1997} is used for representing and reasoning over regions.
RCC describes the possible relations between two regions.
Nutt \cite{nuttTranslationQualitativeSpatial1999} introduced the \emph{topological set constraints} language as a generalization of RCC and reduced reasoning about topological constraints to reasoning in modal propositional logic S4.

The Dimensionally Extended 9-Intersection Model (DE-9IM) \cite{clementiniModelingTopologicalSpatial1994} is a model to describe topological relations that are invariant to translation, rotation, and scaling.
Different relations between two objects are represented in a $3\times3$ capturing the intersections of interior, exterior, and boundary.
This model is used in GIS applications like ArcMap and spatial databases like PostGIS to determine spatial relations between objects \cite{esri2020,ramsey2012}.

The logic programming community has produced many approaches that focused on expressing spatial reasoning directly, rather than by interfacing to a GIS. Most of this work \cite{baryannisTrajectoryCalculusQualitative2018,bhattCLPQSDeclarative2011,pesantReasoningSolidsUsing1999,walegaNonMonotonicSpatialReasoning2017} aims at pushing the boundaries of spatial reasoning to ever more challenging problem categories (e.g. non-monotonic reasoning) rather than on the scalability of the approach or on issues external to reasoning, such as visualisation of the results of the spatial reasoning process. One notable exception is the Space package for SWI-Prolog \cite{hageSpacePackageTight2010}. It also targets efficiency by integrating semantic and spatial indexes in SWI-Prolog. Having both types of indexes in the same system enables advanced query optimization.  
In general, their approach follows the {\entityEvaluation} paradigm, with additional optimization for the NN search.

$\lambda$Prolog(QS) \cite{liLPrologQSFunctional2019} introduces a framework for spatial reasoning within high-order logic programming.
The algebraic semantics of spatial relations is implemented using Constraint Handling Rules.
Using such rules, $\lambda$Prolog(QS) can simplify terms before evaluation the spatial relationships on a numerical level.
The evaluation of the spatial relationships on the numerical level uses a {\entityEvaluation} paradigm.

Geosparql \cite{battleGeoSPARQLEnablingGeospatial2011} introduced a standard for representing and querying spatial data in the Semantic Web.
In Geosparql, spatial features have a special predicate that points to the actual geometry, which is represented as a string.
Whereas Geosparql does not specify the implementation details of the different spatial predicates, the restriction of representing geometries as individual entities implicitly favours a {\entityEvaluation} paradigm.

The data model \emph{stRDF}, the query language \emph{stSPARQL}, and the Geospatial DBMS \emph{Strabon} together form a scalable solution to query spatial data in the Semantic Web~\cite{kyzirakosStrabonSemanticGeospatial2012}.
As backend for spatial operations, the \emph{PostGIS} extension of the \emph{Postgres} DBMS is used.
The paper does not discuss the optimization using a {\relationEvaluation} paradigm and so, to the best of our knowledge, only the {\entityEvaluation} paradigm is used.

PelletSpatial \cite{stockerPelletspatialHybridRCC82009}, which is built on top of OWL 2 reasoner Pellet, implements two Region Connection Calculus (RCC) reasoners.
Since geometries are represented as individual entities, a {\entityEvaluation} paradigm is naturally favoured in this setting.
However, specific evaluation strategies for spatial operations are not discussed in detail.

The Geographica 2 benchmark \cite{ioannidisEvaluatingGeospatialRDF2019} evaluates different spatially enabled RDF stores.
The benchmark emphasizes the scalability of the systems.

The problem of interfacing logic and databases has been extensively studied by the deductive database (DDB) community. The impedance mismatch of tuple-oriented, top-down evaluation of logic programming and set-oriented bottom-up evaluation has let, already in the early eighties, to the agreement that accessing a relational database from Prolog is unacceptably inefficient. This has sparked research on how to simulate Prolog's top-down evaluation in a bottom up, set-oriented fix-point computation. The magic-set \cite{bancilhonNaiveEvaluationRecursively1986} evaluation was a breakthrough step, followed by ever more sophisticated and efficient implementations. Handling recursive programs with function symbols was addressed, among others, in LogicBase \cite{hanLogicBaseDeductiveDatabase1994}. However, most of the research in the DDB community focused on programs without function symbols.

Our work partly confirms the old insight about the inefficiency of tuple-oriented access to a relational (or, in our case, geographic) database. As shown in Section \ref{sec:evaluation}, the purely entity-based approach is indeed orders of magnitude slower. However, we also showed that, in spite of its tuple-oriented evaluation paradigm, logic programming can be a very efficient high-level query language for databases, if used properly. The key point is to let Prolog terms represent entire relations rather than individual database entities. This way, Prolog predicates translate to database queries that exploit the power of set operations and take advantage of all the sophisticated relational and spatial indexing mechanisms of the database.

%% file: conclusion.tex
In this paper, we investigated a new paradigm for spatio-logical reasoning using relations instead of single entities as basic building block.
We have illustrated how this new approach can be used in four different scenarios.
As the scaling evaluation shows, a user can often save one or two orders of magnitude of execution time when switching to the {\relationEvaluation} approach.
In addition, we also discussed the importance of a tight integration with existing GIS software.
To this end, we introduced {\geolog}, a Python Addin for ArcMap.
We have also seen that writing logical rules which operate on relations can become slightly more complicated than the more natural way of writing rules which operate on single entities.
Feature work includes further studies into an automatic conversion of the latter into the former such that a user can benefit from good performance and simpler rules.